\documentclass[twocolumn,showpacs,preprintnumbers,amsmath,amssymb, pr]{revtex4-1}

\usepackage{stmaryrd}
\usepackage{txfonts}
\usepackage{amssymb}
\usepackage{mathrsfs}
\usepackage{graphicx}
\usepackage{dcolumn}
\usepackage{bm}
\usepackage{epsfig}

\begin{document}

\title{Synchronization of One Dimensional Array of Point Josephson Junctions Coupled to a Common Load}

\author{Shi-Zeng Lin\(^{1}\), Xiao Hu\(^{1}\) and Lev Bulaevskii\(^{2}\)}

\affiliation{\(^{1}\)WPI Center for Materials
Nanoarchitectonics, National Institute for Materials Science, Tsukuba 305-0044, Japan\\
\(^{2}\)Los Alamos National Laboratory, Los Alamos, New Mexico 87545, USA}

\date{\today}


\begin{abstract}
We study the synchronization in a one dimensional array of point Josephson junctions coupled to a common capacitor, which establishes a long-range interaction between junctions and synchronizes them. The stability diagram of synchronization in a noise-free system is obtained. The current when junctions transform from resistive state into zero-voltage state, is then calculated and its dependence on the shunt parameters and the dissipation of junctions is revealed. In the presence of thermal noise, the synchronized oscillations are destroyed at a critical temperature and the system undergoes a continuous phase transition of desynchronization. A possible stability diagram of the synchronized oscillations with respect to thermal noise, current, dissipations and shunt capacitance is then constructed. Finally we investigate the dynamic relaxation from random oscillations into synchronized state. The relaxation time increases with the system size and temperature, but is reduced by the shunt capacitor.

\end{abstract}

\maketitle

\section{Introduction}
Josephson junctions are building elements of many electronic and electromagnetic
devices as well as a candidate for quantum computers\cite{BaroneBook}. In practical applications, one usually integrates large arrays of junctions on a chip to enhance the performance, thus coherent operations in these junctions are crucial. The synchronization between junctions can be realized by coupling them to a common resonator, most frequently through electromagnetic coupling. The common resonator establishes long-range interaction between junctions, which then synchronizes them under appropriate condition. The junctions arrays have become an extremely important playground to understand the synchronization mechanism for large population of nonlinear oscillators, partially because of the relatively easy experimental realization.\cite{Hadley88,Jain84,Wiesenfeld96,Durala99, Filatrella00,Grib02,Grib06, Filatrella07,Madsen08}.

The successful observations of coherent emission from cuprate superconductors renew
the interests in understanding the synchronization of arrays of Josephson junctions
\cite{Bae07,Ozyuzer07,kadowaki08,Wang09,Wang10,Tsujimoto10,Krasnov10}. Cuprate superconductors, such as $\rm{Bi_2Sr_2CaCu_2O_{8+\delta}}$ (BSCCO),
are a natural realization of a stack of Josephson junctions of atomic thickness \cite{Kleiner92,Kleiner94},
known as intrinsic Josephson junctions (IJJs). Because of the large supercoducting energy gap,
these build-in Josephson junctions can be operated at frequencies in the terahertz region,
where the electromagnetic waves have wide applications \cite{Ferguson02,Tonouchi07}.

Radiation from IJJs occurs in the resistive state. Such a state is reached
by increasing the bias current above the Josephson critical current and then diminishing it down to the voltage
$V$ corresponding to the target frequency according to the Josephson relation $\omega=2eV/\hbar $.
The resistive state is preserved down to the retrapping current below which the system undergoes transition into the zero-voltage state. Such a procedure is possible because the resistivity of IJJs is very large, i.e. junctions are strongly underdamped. Thus the hysteretic behavior allows us, in principle,  to reach a quite low voltage of the order of that corresponding to the Josephson frequency ($\sim 0.1\rm{THz}$ for BSCCO). In the resistive state, Josephson plasma of composite oscillations of Cooper pairs and electromagnetic waves is excited. If the plasma oscillations are in-phase, then the total radiation power is proportional to the number of junctions squared. Below a threshold current called retrapping current, the resistive state becomes unstable and the system switches into zero-voltage one. Important questions to be addressed are:
\begin{enumerate}
  \item What is the retrapping current in the array of point junctions and how a shunt affects it.
  \item In what parameter region of junction and shunt where oscillations of junctions remain synchronized in the resistive state.
\end{enumerate}

The stability of synchronized oscillations depends crucially on interaction between junctions. The junctions in cuprate superconductors interact
with each other through nearest neighbor coupling, either inductive or capacitive.
These short-range interaction however is insufficient to establish a global phase coherence \cite{Hong05,Mori10}.
There are two methods to achieve global synchronization by coupling all junctions to a common resonator.

In the first approach, the cavity formed by the superconductors single crystal plays a role of the resonator \cite{Ozyuzer07,kadowaki08}. The synchronization is realized by the excitation of cavity mode in the crystal\cite{szlin08b,Koshelev08b}. Alternatively, the synchronization can be achieved by the radiation fields\cite{Bulaevskii07} and/or by a shunted circuit \cite{Martin10}.
The synchronization by a shunted circuit attracts considerable interests, because it can be implemented easily.

Real junctions involve thermal noise, especially for those in high-$T_c$ superconductors. Generally, one expects thermal noise broaden the linewidth of oscillating spectrum, or even destroys the coherence. It is preferable to have robust coherent oscillation against noise. To this end, it is important to know how thermal fluctuations destroy the synchronization.

The dynamical process of building up the synchronization is also important for both applications and theoretical understandings. For an initial condition that is very close to the fully synchronized state, the relaxation to synchronized state can be analyzed based on the standard local stability analysis\cite{Bulaevskii07,Lin09c,Koshelev10a}. However, for a complete random initial state, the dynamic process is highly nontrivial. The system may even not relax into the synchronized state. Two questions naturally arise: how to reach the synchronized state in a controlled way and what is the relaxation time?

In this paper, we consider a one dimensional array of point Josephson junctions
coupled to a common circuit. First we provide analytical
and numerical study on the stability of the synchronized state and map out the stability phase diagram.
Based on the diagram, we derive the dependence of the retrapping current on the shunt circuit.
Then we introduce thermal noise into the system and describe the effect on the synchronization. Mean-field critical behaviors are identified at the desynchronization transition, i.e. transition from the the synchronized state to the state with random or partially random oscillations. We reveal the dependence of the transition temperature on
the shunt circuit and bias current. Based on these results, a possible stability diagram of the synchronized oscillations is constructed taking thermal noise into account.
Finally, we study the relaxation dynamics starting from a disordered state, where junctions oscillate randomly.

The remaining part of the paper is organized as follows.
In Sec. II, we introduce the model. In Sec. III, we perform stability analysis of the synchronized oscillations
both numerically and analytically. In Sec. IV, We study the desynchronization transition of the coherent state and obtain the corresponding transition temperature. In Sec. V, we study the dynamic relaxation from disordered
initial state into the synchronized state. The paper is concluded by a short summary.

\begin{figure}[t]
\psfig{figure=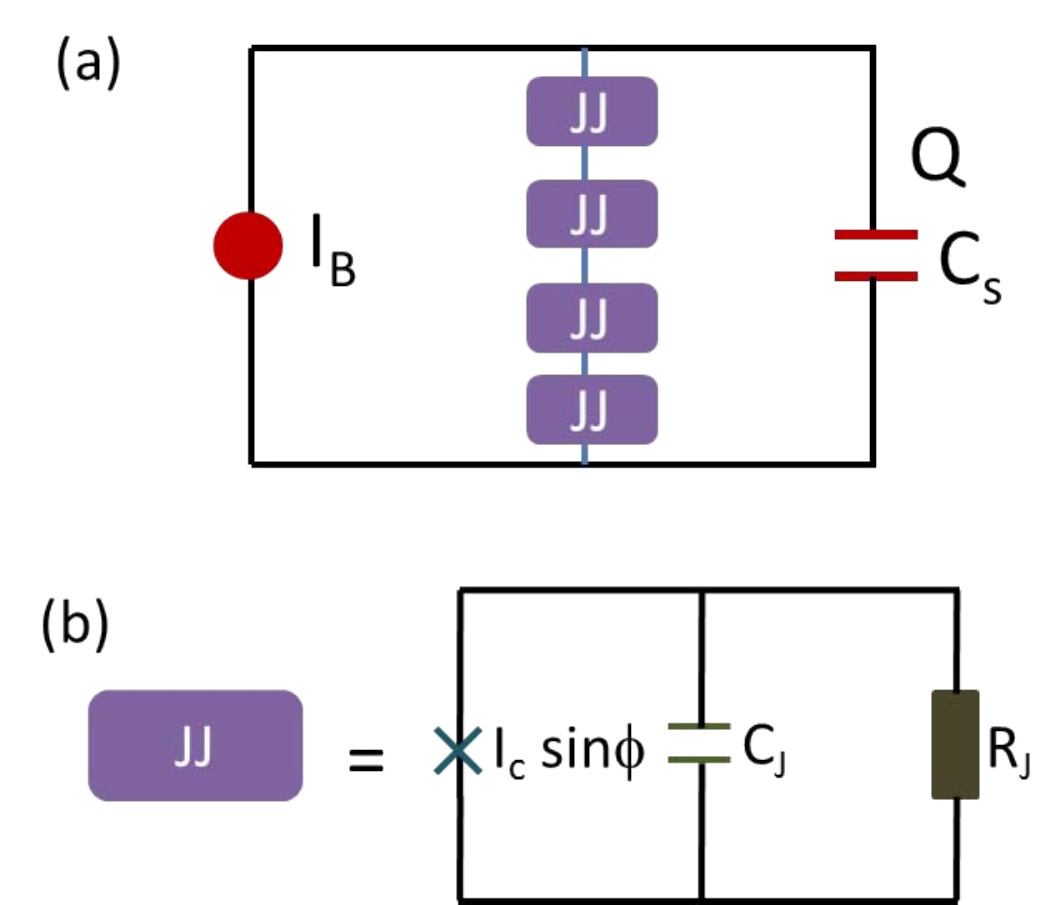,width=\columnwidth} \caption{\label{f0}(Color online). (a) Schematic view of an array of Josephson junctions shunted with a capacitor. The junctions are biased by a dc current $I_B$. (b) The Josephson junction is modeled as a shunt circuit of a capacitor, a resistor and a nonlinear Josephson current.}
\end{figure}

\section{Model}
Arrays of Josephson junctions coupled to a common load have been extensively studied decades ago
\cite{Jain84,Filatrella00,Filatrella07}, not only for their importance for the application
in electronic device, but also as a fruitful platform to understand the underlying synchronization mechanism.
These models although are less transparent than the well known Kuramoto model \cite{Acebron05},
can be realized experimentally\cite{Barbara99,Song09} much easier than the Kuramoto model.
The latter has been realized experimentally only very recently \cite{Kiss02}, long after its proposal.
Some specific configuration of the array, such as one dimensional array of Josephson junctions
shunted by a serial RLC circuit, can be mapped into the Kuramoto model \cite{Wiesenfeld96}.

We consider a stack of IJJs with lateral sizes of order of several micrometers. This geometry of junctions is an alternative route to strong emissions\cite{Bulaevskii07} and has attracted lots of attention recently. In this case, the variation of superconductivity phase in the lateral direction is small and the junction can be approximated as a point junction. The inductive coupling\cite{Bulaevskii91,Sakai93,Bulaevskii94} between junctions then vanishes under this approximation. Meanwhile, the capacitive coupling\cite{Koyama96,Machida99} between junctions is weak and short-range, thus it can be neglected in comparison with the long-range interaction mediated by the shunt circuit. Under these simplifications, a stack of IJJs reduce to a serial array of point junctions.

We study a serial array of point Josephson junctions shunted  by a lumped $C$ circuit, which is
shown schematically in Fig. \ref{f0}. Each junction is modeled as resistively and capacitively shunted circuit. The total current across the junction is
\begin{equation}\label{eq0}
I_J=I_c\text{sin$\phi $}_k+\frac{\hbar }{2e R_J}\dot{\phi }_k+\frac{\hbar }{2e }C_J\ddot{\phi }_k,
\end{equation}
where $V=\frac{\hbar }{2e}\dot{\phi }_k$ is the voltage of the junction according to the ac
Josephson relation. Here $\phi_k$ is the gauge invariant superconductivity phase difference of $k$-th junction, and $R_J$, $C_J$ and $I_c$ are the resistance, capacitance and critical current of the junction respectively.
Using the Kirchhoff's loop law, we obtain the equation of motion
\begin{equation}\label{eq1}
I_B=\dot{Q}+I_c\text{sin$\phi $}_k+\frac{\hbar }{2e R_J}\dot{\phi }_k+\frac{\hbar }{2e }C_J\ddot{\phi }_k+I^n_k,
\end{equation}
\begin{equation}\label{eq2}
V=\frac{\hbar }{2e}\sum_k^N \dot{\phi }_k=\frac{Q}{C_s},
\end{equation}
where $Q$ is the charge on the shunted capacitance, $C_s$ is the shunted capacitance and $I_B$ is the bias DC
current. We have introduced the Nyquist noise (white noise) current $I^n_k$,
\begin{equation}\label{eq3}
\langle I_k^n\rangle =0, \ \ \ \langle I^n_k(t) I^n_{k'}(t')\rangle=(4k_B T/R_J)\delta(t-t')\delta(k-k'),
\end{equation}
where $k_B$ is the Boltzmann constant and $T$ is the temperature. We have also assumed that junctions are identical. In the presence of the common circuit, small spread in the junction's parameters will not destroy the coherent oscillations.

We will use dimensionless quantities in the following calculations. The time is in units of Josephson plasma frequency $\omega_p=\sqrt{2e  I_c/ \hbar  C_J}$, current in units of $I_c$, capacitance in units of $C_J$, resistance in units of $R_J$. We then arrive at the dimensionless version of Eqs. (\ref{eq1}), (\ref{eq2}) and (\ref{eq3})
\begin{equation}\label{eq4}
I_B=\dot{Q}+\text{sin$\phi $}_k+\beta\dot{\phi }_k+\ddot{\phi }_k+I^n_k,
\end{equation}
\begin{equation}\label{eq5}
V=\sum_k^N \dot{\phi }_k=\frac{Q}{C_s},
\end{equation}
\begin{equation}\label{eq6}
\langle I^n_k(t) I^n_{k'}(t')\rangle=2 \beta T \delta(t-t')\delta(k-k'),
\end{equation}
where $1/\beta^2$ with $\beta={\sqrt{\hbar  }}/({ R_J \sqrt{2e I_c C_J}})$ is the McCumber number
which determines the hysteretic behavior of junctions. Upon increasing the bias current, the system remains zero-voltage
until the bias current exceeds the critical current. Then the junctions switch into resistive state.
The system keeps resistive even when the bias current is reduced below the critical current for a junction with small $\beta$\cite{HuReview09}. In the resistive state, the superconductivity phase $\phi_k$ is rotating accompanied by small oscillations. The angular velocity of the rotation for $\phi_k$ is the same for all junctions determined by the voltage, but the phase $\phi_k$ may vary from junction to junction. We will consider the synchronization of the phase of the junction arrays in the following.

In this model, all junctions are coupled to the capacitor $C_s$, which establishes mutual
interaction among all junctions. Thus the effective dimensionality of the system is infinite, which is
crucial for the synchronization \cite{Hong05,Mori10}. Another consequence of this mean-field behavior
is the permutation symmetry. i.e., all junctions are biased by the same external current and the current in the shunt circuit. The exchange of any pair of junctions in the circuit does not change the topology of the circuit. If configuration $(\phi_1, \phi_2, ..., \phi_i, ...,\phi_j, ...,\phi_N)$ is a solution to Eqs. (\ref{eq4}), (\ref{eq5}) and (\ref{eq6}), then $(\phi_1, \phi_2, ..., \phi_j, ...,\phi_i, ...,\phi_N)$ is also a solution. This symmetry greatly simplifies the stability analysis as will be shown below.

Apparently, Eqs. (\ref{eq4}), (\ref{eq5}) and (\ref{eq6}) always have a trivial solution that all junctions oscillate out of phase, and the dynamics of each junction is independent because the current in the shunt circuit vanishes. However suppose at some instance, a small population of junctions oscillate with the same phase, then the capacitance $C_s$ acquires energy, which is proportional to the number of in-phase junctions squared. Now the capacitance is able to attract more junctions to oscillate at its phase and in turn its energy increases further. This is a positive feedback process with explosive increases of energy in the capacitance and an avalanche of junctions oscillating coherently. Therefore we expect in certain parameter space, the out-of-phase oscillations lose stability and synchronization sets in. The qualitative picture will be elaborated in subsequent sections.

\section{Stability of the Synchronized State}

In this section, we analyze the local stability of the coherent oscillations of all junctions in the absence of
thermal fluctuations, $T=0$. The local stability is determined by the dynamics of the system in the vicinity
of the trajectory of the uniform solution. We consider the uniform oscillations $\phi_k=\phi_0$, where
 \begin{equation}\label{eq7}
(N C_s + 1){\ddot \phi _0} + \beta {\dot \phi _0} + \sin {\phi _0} = {I_B},
\end{equation}
with $N$ being the number of junctions. The junction coupling strength is enhanced by a factor of $N$,
in accordance with the typical behavior in the mean-field theory.
We then add small perturbations $\delta_k$ to the uniform solution and determine
the time evolution of the perturbations. The equations for the perturbations read
\begin{equation}\label{eq8}
{\ddot \delta _k} + \beta {\dot \delta _k} + \cos ({\phi _0}){\delta _k} + C_s\sum\limits_{i = 1}^N {{{\ddot \delta }_i}}  = 0.
\end{equation}
The permutation symmetry between junctions allows us to decouple Eqs. (\ref{eq8}) by introducing
the quantities $\Delta_k=\delta_{k+1}-\delta_{k}$ and
$\sigma  = \frac{1}{N}\sum\limits_{i = 1}^N {{\delta _k}}$. We obtain equations for $\Delta_k=\Delta$ and $\sigma$:
\begin{equation}\label{eq9}
\ddot \Delta  + \beta \dot \Delta  + \cos ({\phi _0})\Delta  = 0,
\end{equation}
\begin{equation}\label{eq10}
\ddot \sigma  + \beta \dot \sigma  + \cos ({\phi _0})\sigma  + C_s N\ddot \sigma  = 0.
\end{equation}
If $\Delta$ diverges with time, the uniform solution becomes
unstable. On the other hand, if $\sigma$ diverges while $\Delta$ decays with time, the synchronization is kept and the system transits into another synchronized state if it exists.
We are interested in the coherent oscillation, and we will only focus on Eq. (\ref{eq9}) in the later analysis.
We will solve Eq. (\ref{eq9}) for weak oscillations both analytically and numerically based on the Floquet theorem.

\begin{figure}[t]
\psfig{figure=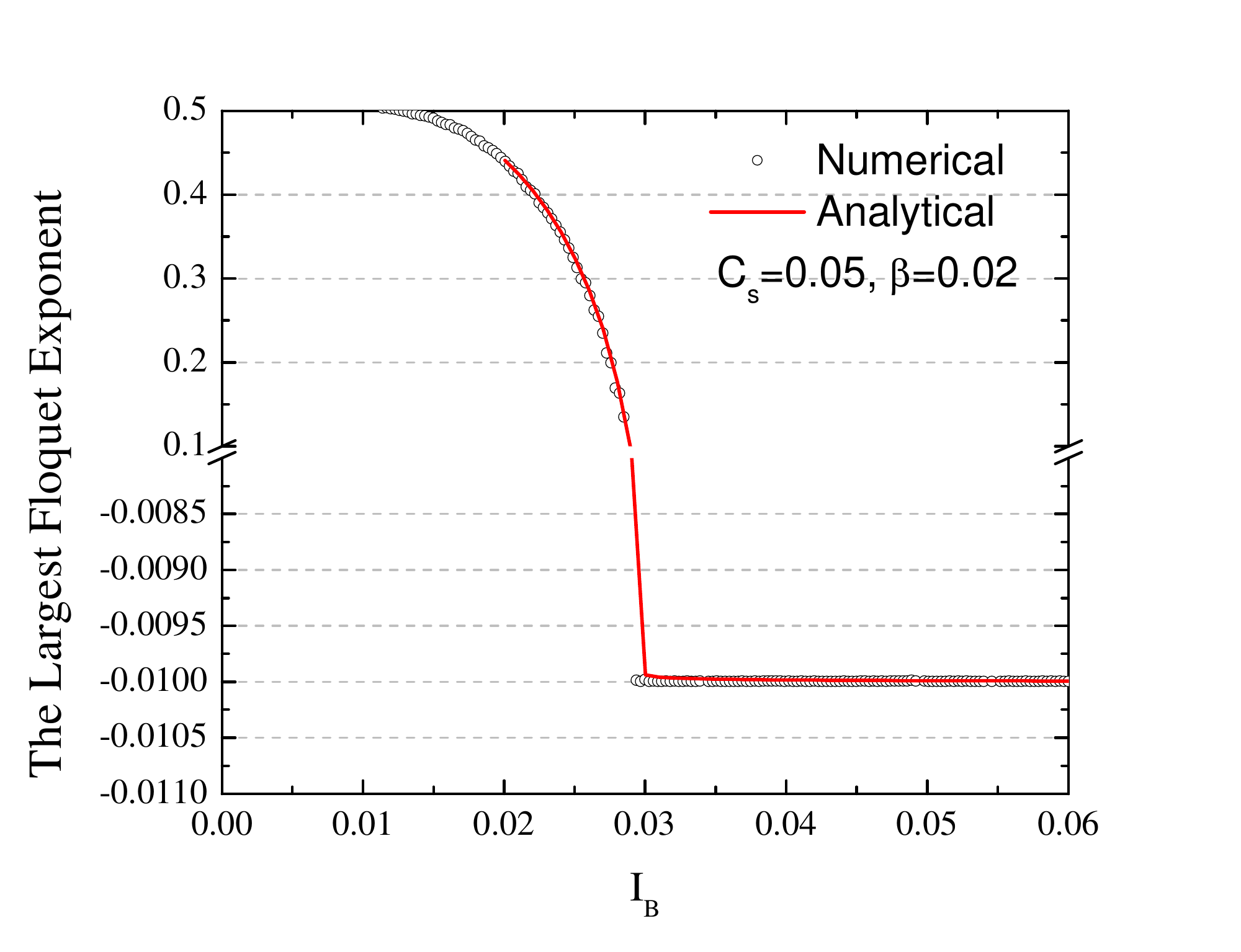,width=\columnwidth} \caption{\label{f1}(Color online). The largest Floquet exponent calculated by $\det \mathbf{D}=0$ with $\mathbf{D}$ given by Eq. (\ref{eq15}) [red line], and by numerical calculation using the Floquet theory in Eq. (21) [symbols]. For the exponent smaller than 0, the uniform oscillations are stable.}
\end{figure}

\subsection{Analytical treatment}

We consider the region where the amplitude of Josephson oscillation is small.
The solution of Eq. (\ref{eq7}) in  linear approximation can be written as
\begin{equation}\label{eq11}
\phi_0=\omega  t+A \exp (i \omega  t)
\end{equation}
with
\begin{equation}\label{eq11aa}
A=\frac{i}{-(C_s N+1)\omega ^2+i \beta \omega}\ll1.
\end{equation}
The frequency $\omega$ is
determined by the DC current conservation
\begin{equation}\label{eq11b}
I_B=\beta  \omega +\text{Re}[A]/2.
\end{equation}
Substituting Eq. (\ref{eq11}) into Eq. (\ref{eq9}), we get the equation for $\Delta$
\begin{equation}\label{eq12}
\ddot{\Delta }+\beta  \dot{\Delta }+\left[\frac{1}{2}\left(e^{i \omega  t}+e^{-i \omega  t}\right)-\frac{1}{2i}\left(e^{2 i \omega  t}-1\right) A \right]\Delta =0.
\end{equation}
The coupling of perturbations to the oscillation $\exp(i\omega t)$ induces higher frequency harmonics.
The general solution for $\Delta$ is
\begin{equation}\label{eq13}
\Delta =e^{-i \Omega  t}\sum _{k=-\infty}^{+\infty} a_k e^{i k \omega  t}.
\end{equation}
The stability is determined by the spectrum of perturbations $\Omega$. In the framework of the Floquet
theory\cite{MagnusBook}, we call $\text{Im} (\Omega)$ as the Floquet exponent. The uniform solution is stable if and only if the largest Floquet exponent is negative, i.e. $\text{Im} (\Omega)<0$. One may easily identify $\text{Im} (\Omega)$ as a relaxation time and $\text{Re} (\Omega)$ as an energy gap of perturbations.

To obtain $\Omega$, we plug Eq. (\ref{eq13}) into Eq. (\ref{eq12}) and compare each frequency component. Then we have the following linear equations for the coefficients $a_k$
\begin{equation}\label{eq14}
-(k \omega -\Omega )^2a_{k}+i (k \omega -\Omega ) \beta  a_{k}+\frac{1}{2}\left(a_{k-1}+a_{k+1}\right)-\frac{A}{2i}\left(a_{k-2}-a_k\right)=0
\end{equation}
The existence of nonzero solution of $a_k$ requires the determinant of the coefficient matrix vanishes, $\det \mathbf{D}=0$ with
\begin{equation}\label{eq15}
\mathbf{D}=\left(
\begin{array}{ccccccc}
 \text{...} & \text{..} & \text{..} & \text{..} & \text{..} & \text{..} & \text{..} \\
 -\frac{A}{2i} & \frac{1}{2} & c_{-1} & \frac{1}{2} & 0 & 0 & 0 \\
 0 & -\frac{A}{2i} & \frac{1}{2} & c_0 & \frac{1}{2} & 0 & 0 \\
 0 & 0 & -\frac{A}{2i} & \frac{1}{2} & c_1 & \frac{1}{2} & 0 \\
 0 & 0 & 0 & -\frac{A}{2i} & \frac{1}{2} & c_2 & \frac{1}{2}\\
  \text{...} & \text{..} & \text{..} & \text{..} & \text{..} & \text{..} & \text{..}
\end{array}
\right)
\end{equation}
and $c_k= -(k \omega -\Omega )^2+i (k \omega -\Omega ) \beta  +{A}/({2i})$. The solution gives the spectrum of the perturbation.

In the region of $\omega\gg1$, the frequency modes with $k=0, \pm 1$ are dominant and higher harmonics may be
truncated. We obtain a second order equation for $\Omega$
\begin{equation}\label{eq16}
\Omega ^2+i \beta  \Omega  =\frac{A}{2i}+\frac{1}{2\omega ^2}=\frac{C_s N}{(C_s N+1)}\frac{1}{2\omega ^2},
\end{equation}
with solutions
\begin{equation}\label{eq17}
\Omega =\frac{-i \beta \pm \sqrt{-\beta ^2+\frac{2C_s N}{(C_s N+1)}\frac{1}{\omega ^2}}}{2}.
\end{equation}
We see that the uniform oscillations are always stable for non-zero $C_s$ in the region of $\omega\gg1$.
In the limit $\omega\approx I_B/\beta\rightarrow \infty$, the largest Floquet exponent approaches zero,
and the solution becomes neutrally stable.

Near the stability boundary where the largest $\text{Im} (\Omega)$ changes sign, one has to keep higher harmonics in Eq. (\ref{eq13}) because $\omega\sim~1$. But for $C_s N\gg1$, one can still use the linear expansion in Eq. (\ref{eq11}). Under these conditions, the stability boundary can be determined by the numerical calculation of $\det \mathbf{D}=0$.

\begin{figure}[t]
\psfig{figure=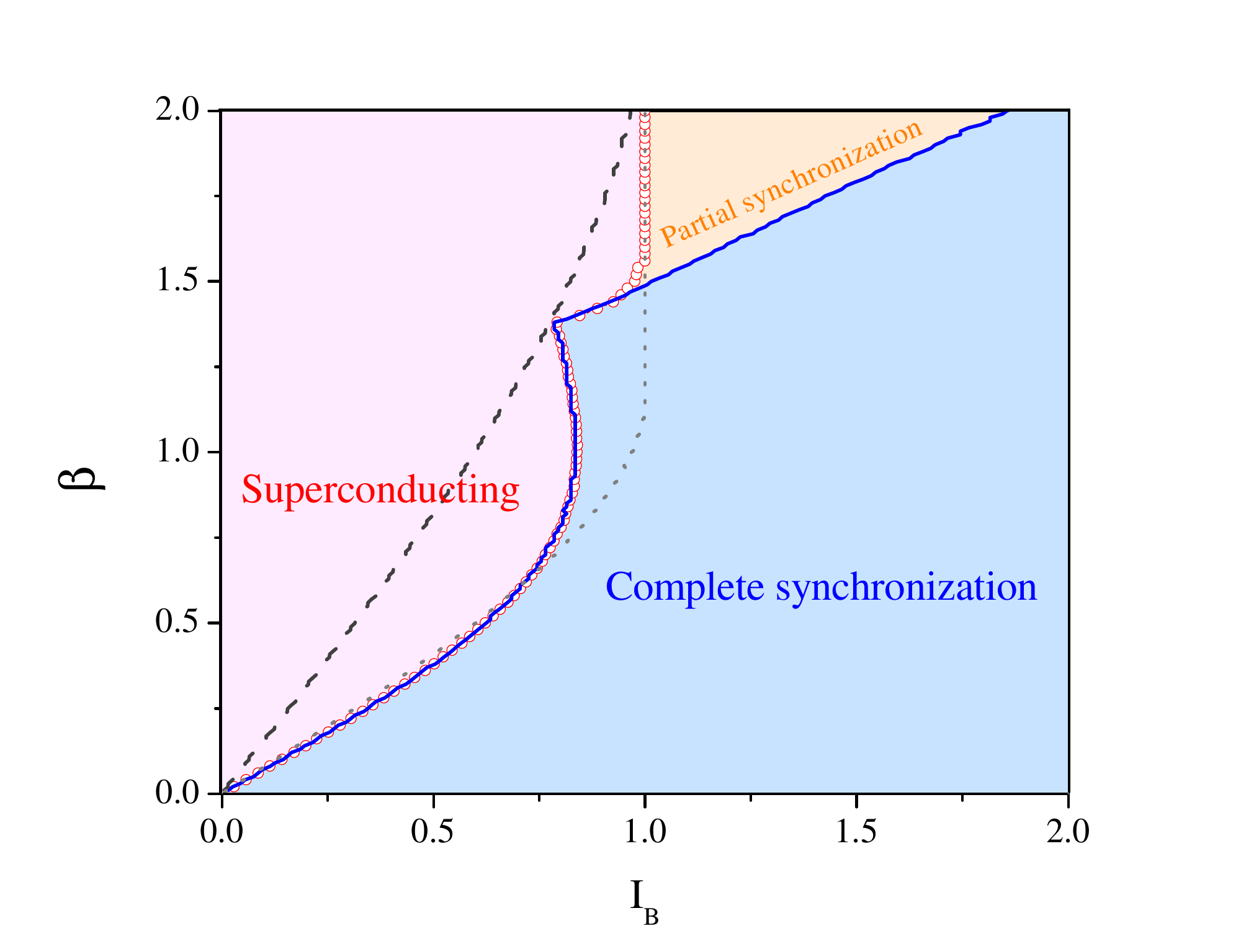,width=\columnwidth} \caption{\label{f2}(Color online).
Stability diagram of the uniform solution in the absence thermal fluctuations. Light blue/pink/orange region denotes complete synchronization/zero-voltage state/partial synchronization. The blue line is the stability boundary of the uniform solution calculated by the Floquet theory $I_s$, and the open red circle is the retrapping current $I_r$ determined by direct calculations of Eqs. (\ref{eq4}-\ref{eq6}). The dashed line the retrapping current determined by Eq. (\ref{eq7}) while the dotted line is the retrapping current for a single junction. Here $C_s=3/N$.}
\end{figure}
\begin{figure}[b]
\psfig{figure=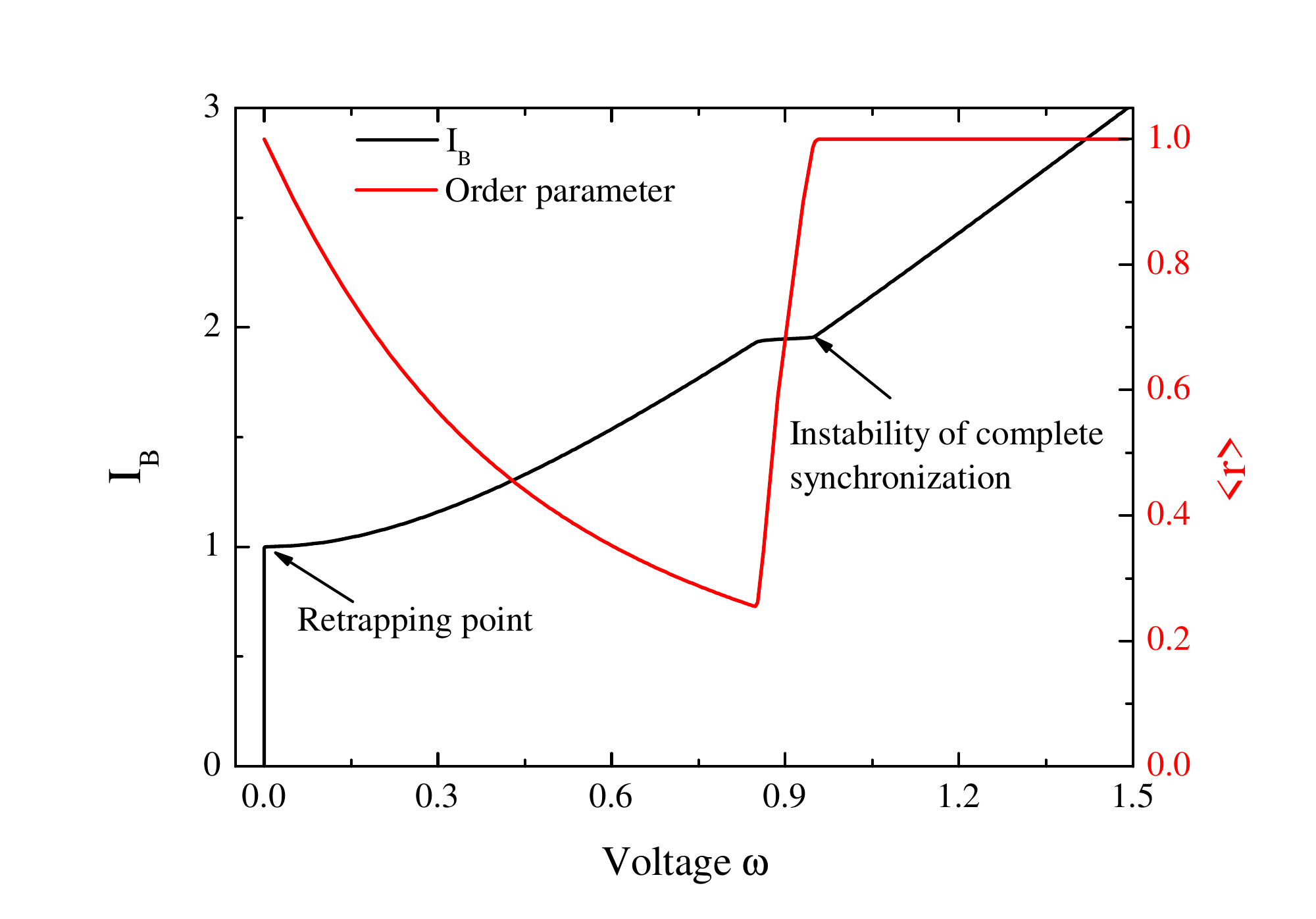,width=\columnwidth} \caption{\label{f3}(Color online). \emph{IV} curve(black) and dependence of the order parameter on the voltage(red) for $\beta=2.0$ and $C_s=3/N$.}
\end{figure}

\subsection{The Floquet theory}

Equation (\ref{eq9}) can also be interpreted as a particle moving in a periodic potential with period $T$. Then we can apply the Floquet theorem (Bloch theorem) to extract the exponents. The solution has the form
 \begin{equation}\label{eq20}
\Delta (t) = \exp ({\lambda _1}t){y_1}(t) + \exp ({\lambda _2}t){y_2}(t),
\end{equation}
where $y_1(t)$ and $y_2(t)$ are periodic functions with period $T$, and the exponents
$\lambda_1$ and $\lambda_2$ follow $\lambda_1+\lambda_2=-\beta$ according to the Floquet theorem.\cite{MagnusBook} When no dissipation is present $\beta=0$, the dynamics is time reversal and $\lambda_1+\lambda_2=0$. When the dissipation is involved, the volume of phase space is shrinking with a rate $\beta$, thus the two exponents follow $\lambda_1+\lambda_2=-\beta$.

The exponents can be computed numerically as follows. We first calculate the trajectory of $\phi_0$ in Eq. (\ref{eq7}). Then we calculate two trajectories of $\Delta_a(t)$ and $\Delta_b(t)$ with two different initial conditions $\Delta_a(t_0)=0$, $\dot\Delta_a(t_0)=1$ and $\Delta_b(t_0)=1$, $\dot\Delta_b(t_0)=0$. These two trajectories obey\cite{MagnusBook}
\begin{equation}\label{eq21}
\left(
\begin{array}{c}
 \Delta _a(t+T),  \Delta _b(t+T)\\
 \dot{\Delta} _a(t+T),  \dot{\Delta} _b(t+T)
\end{array}
\right)=\mathbf{F}(T)\left(
\begin{array}{c}
 \Delta _a(t), \Delta _b(t) \\
 \dot{\Delta} _a(t), \dot{\Delta} _b(t)
\end{array}
\right),
\end{equation}
with $\mathbf{F}$ being a coefficient matrix, which can be evaluated by inverting Eq. (\ref{eq21}) because the trajectories of $\Delta_a$ and $\Delta_b$ are known. $\lambda_1$ and $\lambda_2$ are just the eigenvalues of the matrix $\mathbf{F}$.
We have compared the results obtained by Eq. (\ref{eq21}) and those by analytical calculations. Both methods give the consistent results as shown in Fig. \ref{f1}.

\subsection{Stability diagram}

The stability analysis above does not tell us what the final state is when the uniform solution becomes
unstable. To answer this question, we solve Eq. (\ref{eq4}) and (\ref{eq5}) with $I^n_k=0$ directly
by numerical simulation. The stability diagram then is constructed, and is depicted in
Fig. \ref{f2}. For a sufficient large $I_B$ thus $\omega\gg1$, the uniform solution is stable as described by
Eq. (\ref{eq17}). For a small $\beta$, upon decreasing $I_B$, the uniform oscillations become unstable below the retrapping current $I_r<I_c$ and the system evolves into zero-voltage state. For a large $\beta$, the uniform oscillation loses stability at $I_s>I_c$, where no zero-voltage state is available for the system to go. In this case, the system becomes partially synchronized with a fraction of junctions oscillating in-phase, while the others do out-of-phase oscillation. When $I_B$ is reduced further, the partial synchronization becomes unstable and the system is retrapped into zero-voltage state at $I_B=I_c$.

Let us discuss the transition from the complete synchronization to the partial synchronization
when $\beta$ is large. To characterize the partially synchronized state, we introduce the order parameter which is widely used in literatures\cite{Acebron05}
\begin{equation}\label{eqmelt1}
r(t)\exp[i\theta(t)]=\frac{1}{N}\sum_j^N \exp(i\phi_j).
\end{equation}
Here $r$ is positively defined. We compute the average of $r(t)$
\begin{equation}\label{eqmelt2}
\langle r \rangle=\frac{1}{t_f}\int_{0}^{t_f} dt\ r(t).
\end{equation}
and take $t_f\rightarrow+\infty$.

The \emph{IV} and the corresponding order parameter are shown in Fig. \ref{f3}. When the complete
synchronization becomes unstable, a sharp jump of voltage is observed, associated with decrease
of the order parameter. The reduction of voltage when the system becomes partially synchronized can be understood as
follows. At a given voltage, the shunt capacitor reduces the plasma oscillation amplitude depending on the number of the synchronized junctions, as described by Eq. (\ref{eq11aa}). For the uniform oscillations, the suppression is largest and the DC current induced by the Josephson oscillation is reduced significantly according to Eq. (\ref{eq11b}). For the partial synchronized oscillations, the DC current is larger than that of the uniform oscillations. Therefore when one biases the array with a fixed current, the voltage of the uniform state increases compared with that in partial synchronized state.

\begin{figure}[t]
\psfig{figure=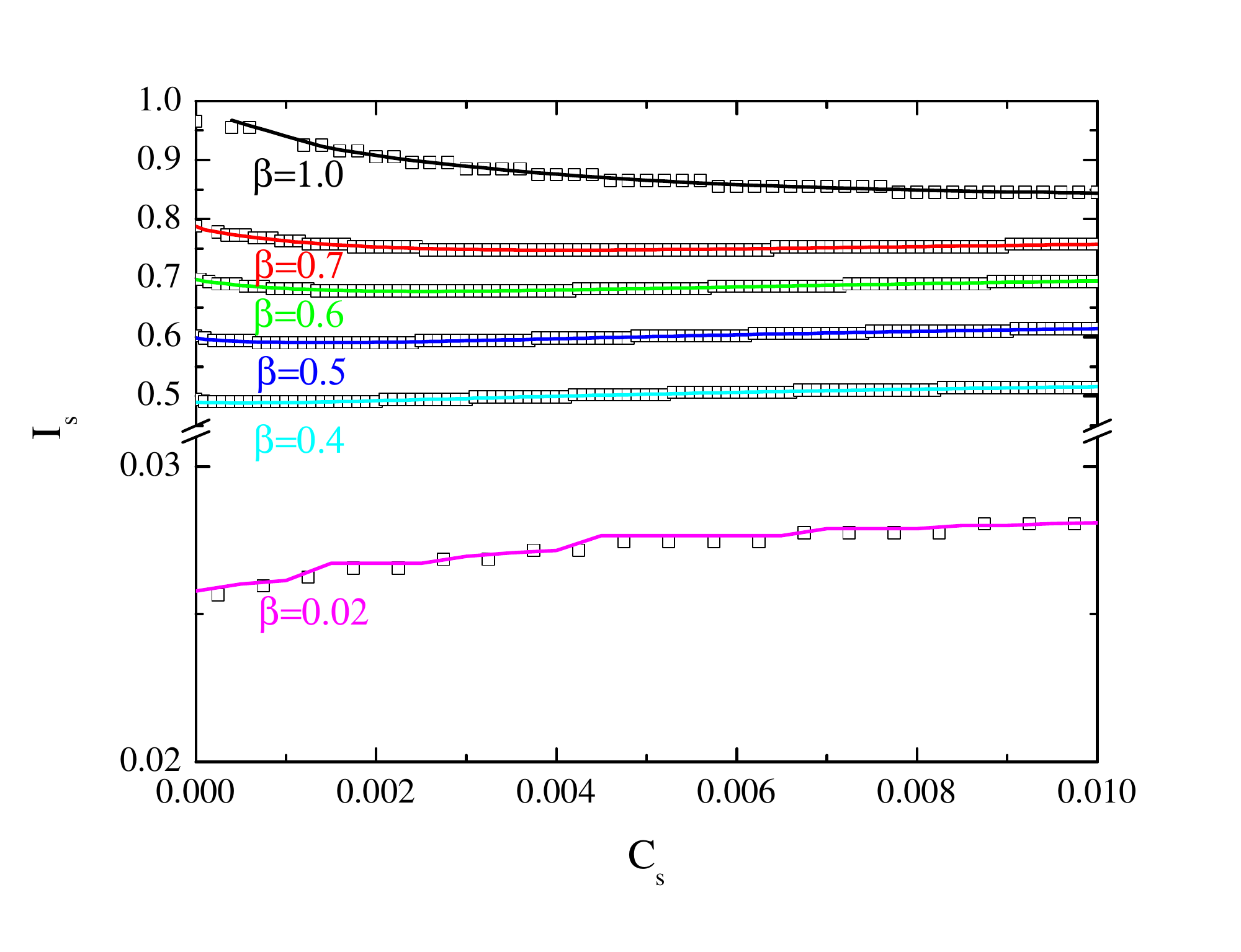,width=\columnwidth} \caption{\label{f4}(Color online). Dependence of the boundary retrapping current $I_r$ on the shunt capacitance for several $\beta$'s. Lines are obtained with the Floquet theory and symbols are direction simulations of Eqs. (\ref{eq4}-\ref{eq6}). The region above the line corresponds to the uniform oscillations while zero-voltage state below the line. Here $N=200$.}
\end{figure}

\subsection{Retrapping current}

According to Eqs. (\ref{eq11aa}), (\ref{eq11b}) the amplitude of Josephson oscillation increases with decreasing $I_B$,
To achieve the strongest oscillation, one would like to know how small current one can achieve in order to support the resistive state.

For a single junction, the dynamics is equivalent to a particle sliding down in the damped inclined washboard potential. It shows hysteretic behavior for small $\beta$, i.e. the system remains resistive even $I_B<I_c$. The system evolves into the superconduction state at a current $I_r>0$, where the input power is insufficient for the phase particle to move in the damped tilted washboard potential. The retrapping current for a weak damping is given by $I_r\approx 1.48 \beta$.\cite{szlin09a} On the other side, the dynamics becomes overdamped for a large $\beta$, and the system comes back to zero-voltage state once $I_B<I_c$. The dependence of $I_r$ on $\beta$ for a single junction is shown by dotted curve in Fig. \ref{f2}.

For junction array shown in Fig. \ref{f0}, if the uniform solution is always stable in the whole current region, the retrapping current will be the same as in a single junction case with an effective
$\beta'=\beta/\sqrt{C_s N+1}$ normalized by the shunt capacitor (dashed line in Fig. \ref{f2}).
In fact, the uniform solution loses stability at $I_s$ (blue line in Fig. \ref{f2}) and the system evolves into the zero-voltage state. Therefore $I_s$ is the genuine retrapping current for the present junctions array, and can be measured experimentally.

\begin{figure}[t]
\psfig{figure=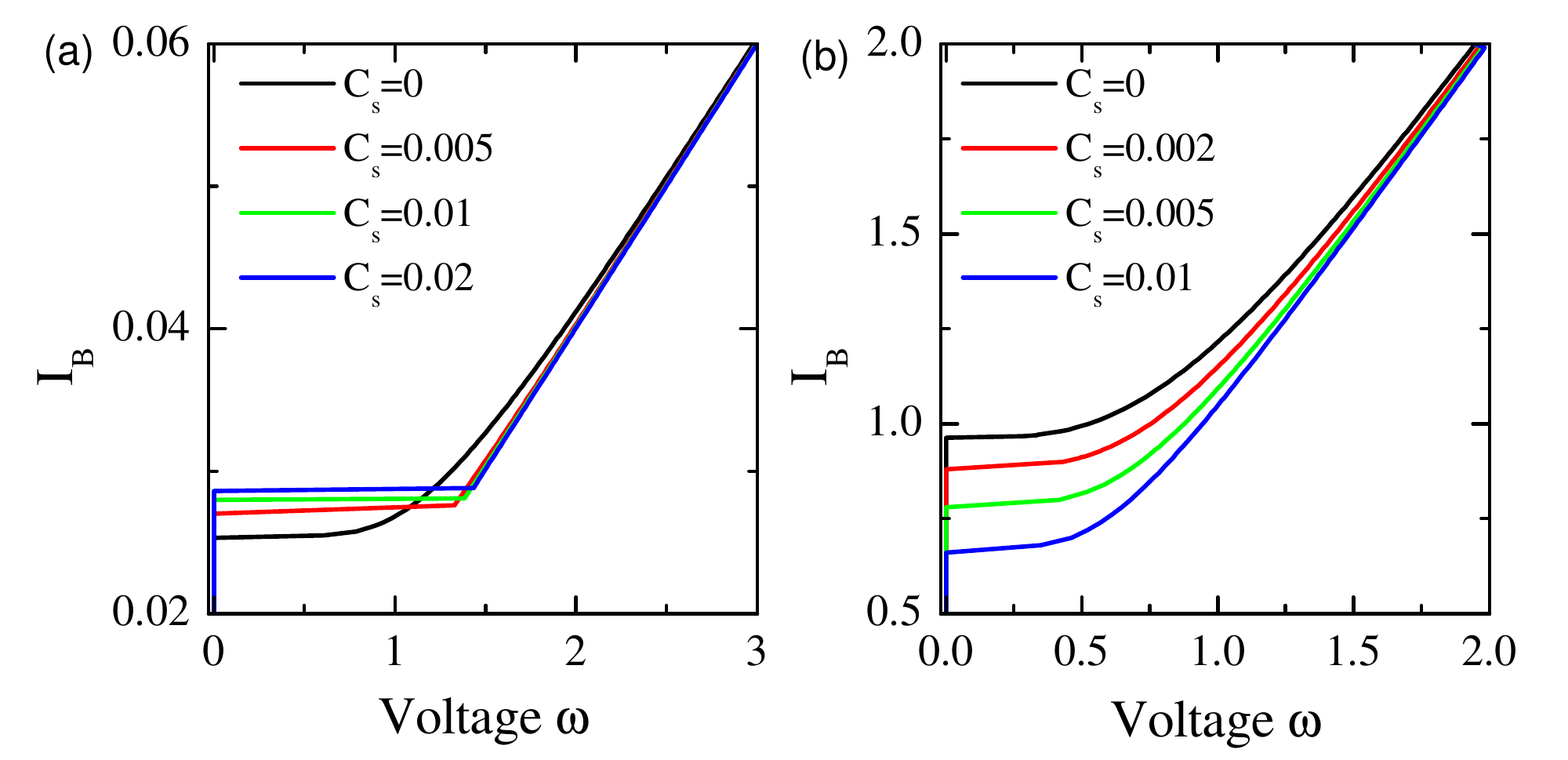,width=\columnwidth} \caption{\label{f6}(Color online). \emph{IV} curves for (a) $\beta=0.02$ and (b) $\beta=1.0$ for several typical values of $C_s$. Here $N=200$.}
\end{figure}

How to decrease $I_s$ or is it possible to shift the stability boundary in Fig. \ref{f2} leftward? One recalls that the shunt capacitor induces interaction between junctions. By increasing the coupling constant $C_s$, one would expect that the stable region enlarges and the stability boundary shift leftward. We study the
dependence of $I_s$ on $C_s$, and the results are presented in Fig. \ref{f4}.
For $\beta\gtrsim 0.5$, the retrapping current decreases with $C_s$, while it increases with $C_s$ for smaller $\beta$. A qualitative picture for this unexpected non-monotonic dependence is as follows.

Equations (\ref{eq7}) and (\ref{eq9}) with $C_s=0$ can also describe the stability of the resistive state for a single junction, where the retrapping current is enhanced by the dissipation as shown by the dotted line in Fig. \ref{f2}. We rewrite Eq. (\ref{eq7}) to a form equivalent to the single junction case by rescaling the time $t\leftarrow \sqrt{N C_s+1}t'$, and with an reduced dissipation $\beta'=\beta/\sqrt{N C_s+1}$. The dynamics of the perturbations  Eq. (\ref{eq9}) then acquire a form
\begin{equation}\label{eqref1}
\frac{1}{N C_s+1}\ddot \Delta  + \beta' \dot \Delta  + \cos ({\phi _0})\Delta  = 0.
\end{equation}
Thus the presence of shunt capacitor reduces both the effective mass of perturbations and damping coefficient. The reduction of the mass makes the system more vulnerable to the oscillatory potential $\cos ({\phi _0})\Delta^2/2$, therefore tends to increase the retrapping current. For the junction array with weak damping, the effect of the reduction of mass dominates and the retrapping current is increased by the shunt capacitor. On the other hand, for the junction array with strong damping, the reduction of the damping by the shunt capacitor dominates and the retrapping current is decreased.

The \emph{IV} curves  for several typical values of $C_s$ are shown in Fig. \ref{f6}. As seen in Fig. \ref{f6} , the \emph{IV} curves with $C_s=0$ deviate from the asymptotic linear behavior $I_B\approx \omega/\beta$ at small $\omega$,  which indicates strong plasma oscillations according to Eq. (\ref{eq11b}). For $\beta=0.02$ the \emph{IV} curves in Fig. \ref{f6}(a) behave differently near the trapping point for $C_s=0$ and for nonzero $C_s$, where in the latter case the \emph{IV} is linear down to the retrapping current.  This linear dependence near the retrapping point with $\omega\sim1$ is due to the suppression of the oscillation amplitude by the shunt capacitor $C_s$. However, for $\beta=1.0$ the \emph{IV} curves remain nonlinear near the retrapping point even for the same $C_s$. For a large $\beta$, the retrapping voltage is much smaller than that of small $\beta$, thus results in stronger plasma oscillations.

\begin{figure}[t]
\psfig{figure=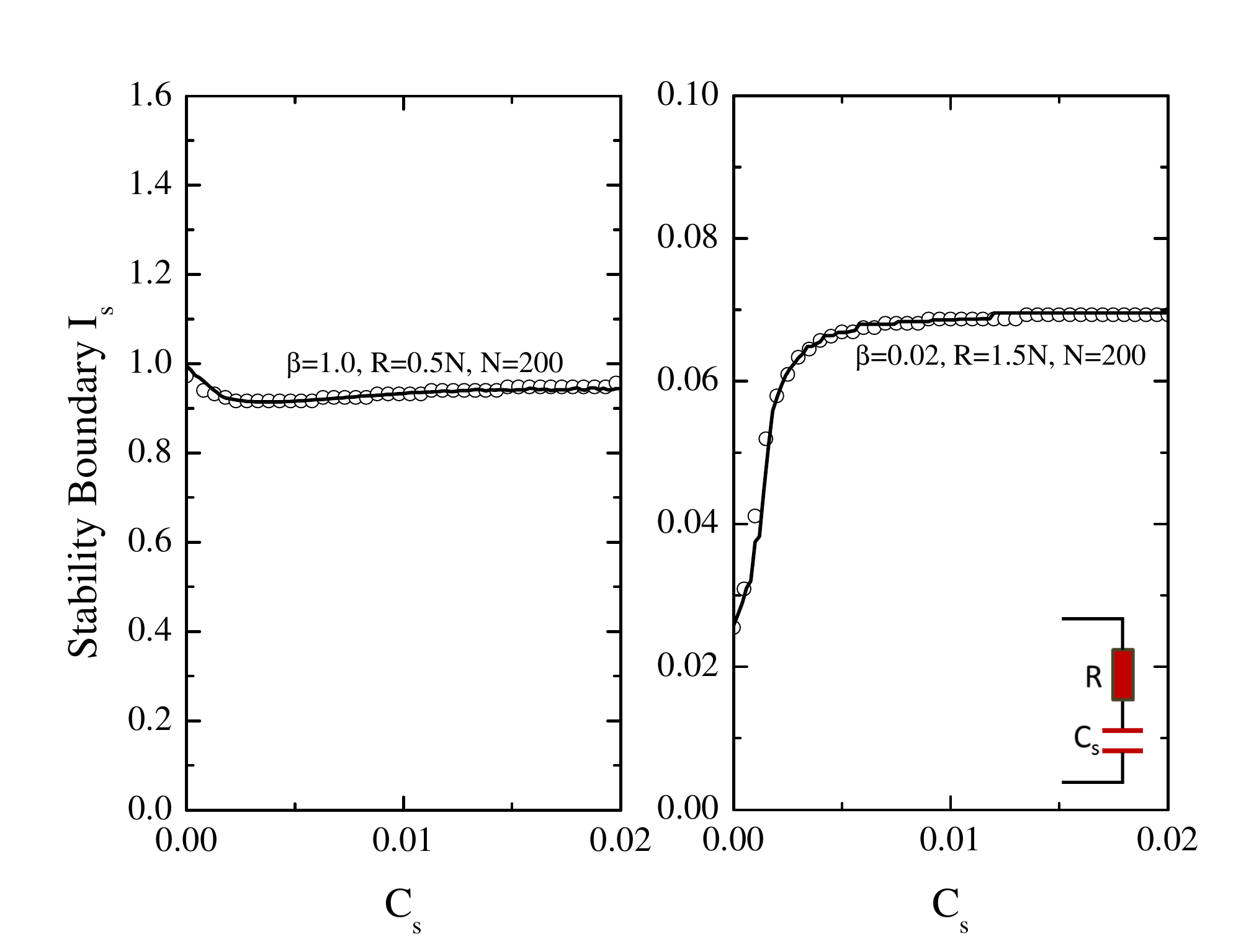,width=\columnwidth} \caption{\label{f7}(Color online).Same as Fig. \ref{f4} but with a shunt RC circuit. Parameters are indicated in the figures.}
\end{figure}
\begin{figure}[t]
\psfig{figure=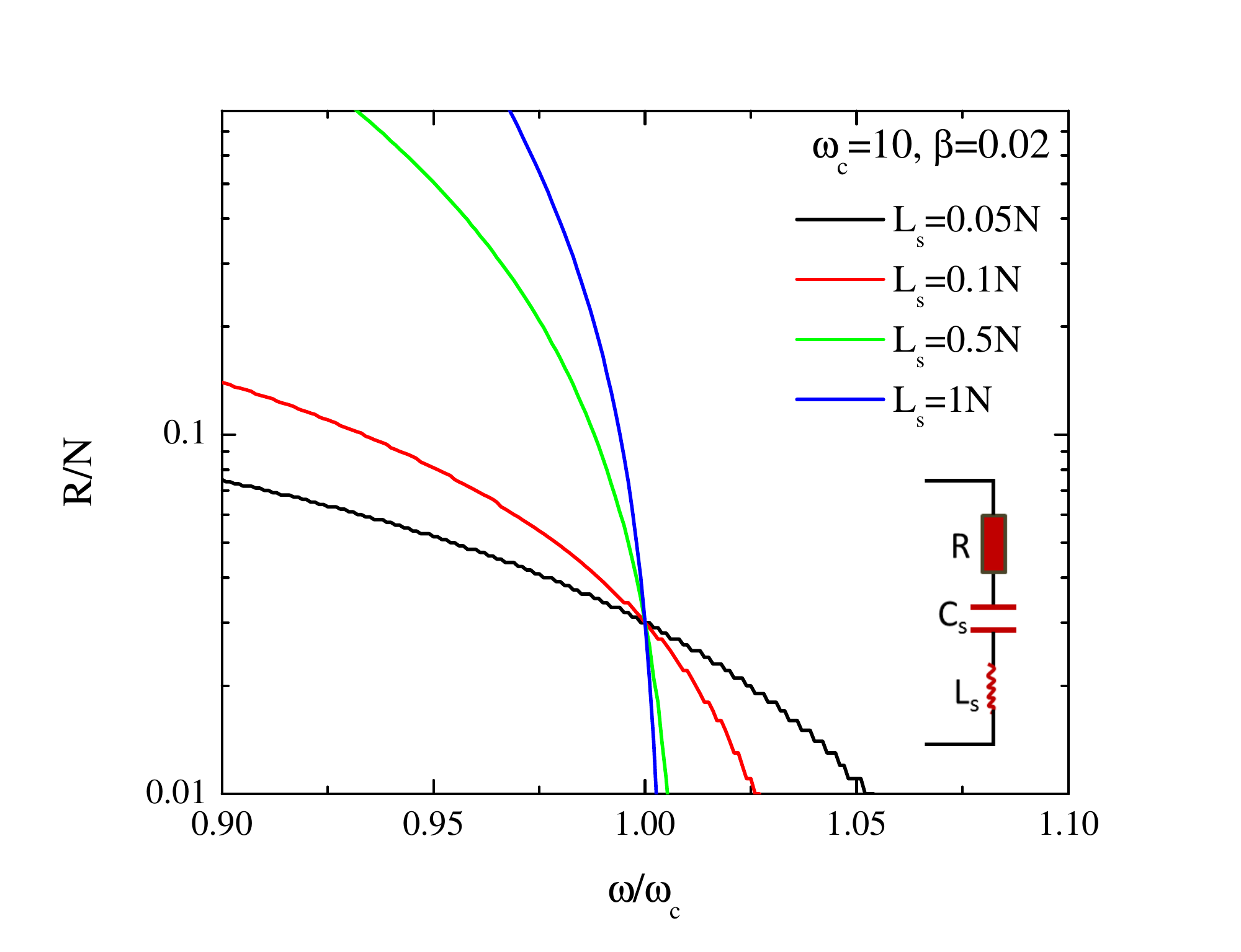,width=\columnwidth} \caption{\label{f7a}(Color online).Stability of the uniform oscillations when an array of Josephson junctions are shunted by an LRC circuit. The region below lines are stable.}
\end{figure}

\subsection{RC circuit}
In real devices, the shunt circuit also carries finite resistance. We perform similar analysis of the stability diagram and the retrapping current, when a resistor with resistance $R$ is serially connected to the capacitor in the load circuit. A new term $\dot{Q} R$ should be added to the right-hand side of Eq. (\ref{eq5}).

The resulting stability diagram is qualitative the same as Fig. \ref{f2}. For a given $R$, the retrapping current increases with $C_s$ for a small $\beta$ while decreases for large $\beta$, as depicted in Fig. \ref{f7}. Comparing Fig. \ref{f7} with Fig. \ref{f4}, we can see that the shunt resistor increases the retrapping current, because the dissipation of the system is increased by the resistor.

\subsection{LRC circuit}
Another interesting case is an array of Josephson junctions shunted by a LRC circuit, which introduces a characteristic frequency $\omega_c=1/\sqrt{L_s C_s}$. The LRC circuit can represent the cavity intrinsically formed by the single crystal of BSCCO \cite{Ozyuzer07,kadowaki08,Tachiki11,Soriano96}.

The stability of the uniform oscillations can be obtained similarly. We consider the case with $\omega\gg1$ and $\beta\ll1$ so the analysis in Sec. III(A) is applicable. The dynamics for small perturbations is still given by Eqs. (\ref{eq9}) and (\ref{eq11}) with a modified amplitude
\begin{equation}\label{eqLRC1}
A=\frac{i}{-\omega ^2+i \beta \omega -\frac{N \omega ^2 }{L_s\left(\omega _c^2- \omega ^2\right) +\omega  i R}}
\end{equation}
The stability is determined by Eq. (\ref{eq16}) and the results are shown in Fig. \ref{f7a}. When $\omega\ll\omega_c$, the LRC circuit behaves as a RC circuit, which always stabilizes the uniform solution for a small $R$ as given by Eq. (\ref{eq17}). On the other hand, for $\omega\gg\omega_c$, the LRC circuit behaves as a LR circuit which makes the uniform solution unstable for a large $L_s$. When $\omega\sim\omega_c$, the stability depends on the quality factor $R$ of the LRC circuit. Small quality factor (large $R$) makes the synchronization difficult.

\begin{figure}[t]
\psfig{figure=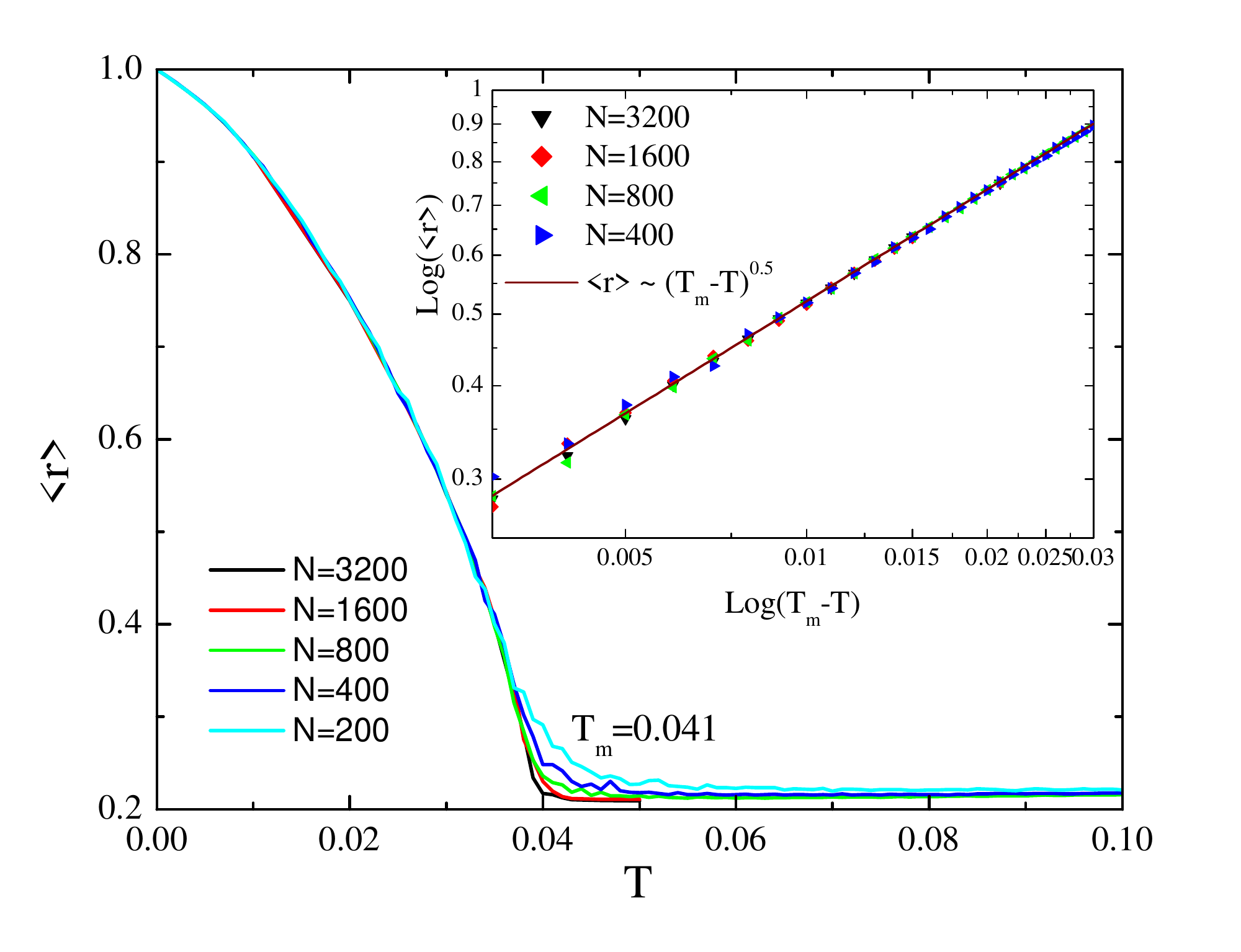,width=\columnwidth} \caption{\label{f8}(Color online). Dependence of the order parameter on temperature with different system sizes. Inset is a double-logarithm plot of the reduce temperature $T_m-T$ and order parameter. Here $I_B=1.5$, $\beta=1.0$ and $C_s=3.0/N$.}
\end{figure}
\begin{figure}[b]
\psfig{figure=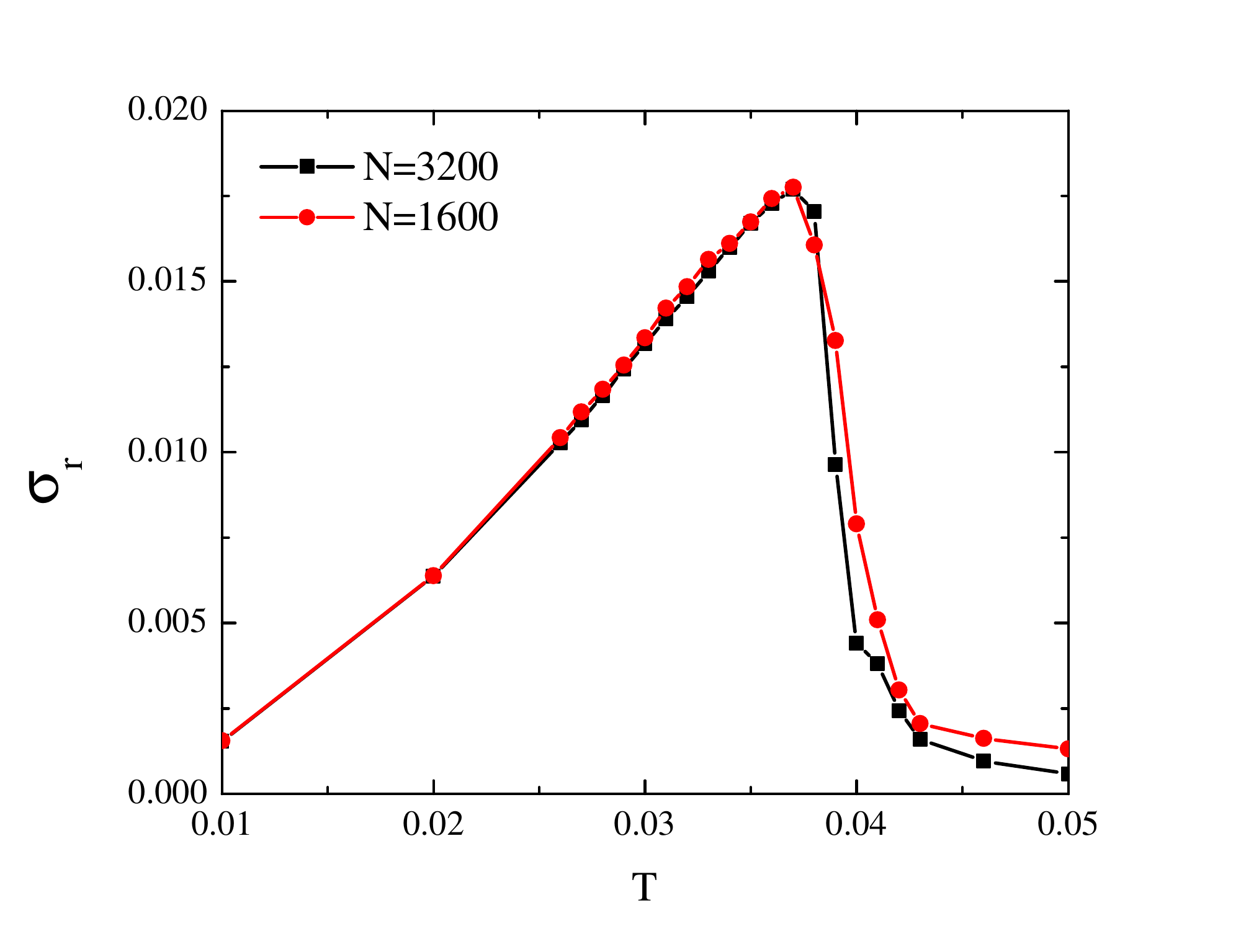,width=\columnwidth} \caption{\label{f9}(Color online). Dependence of the fluctuations $\sigma_r$ on the temperature. Here $I_B=1.5$, $\beta=1.0$ and $C_s=3.0/N$.}
\end{figure}

\section{Effect of Thermal noise}

Real circuits inevitably involve noise because of resistivity caused by quasiparticles. This leads to diffusive dynamics in the phase space and destroys the synchronization at a certain critical point. To study the effect of noise, knowledge of attractors in the phase space is necessary. An attractor attracts trajectories nearby and the volume of phase space that the attractor attracts defines the basin of attraction of the attractor. The phase space is covered by basins of attraction and the boundary between basin of attraction is called separatrix. For a small $\beta$, we have already identified two attractors with one being the zero-voltage state, and the other uniform-oscillation state. For a large $\beta$, additional attractor with partial synchronous oscillations appears.

\begin{figure*}[t]
\psfig{figure=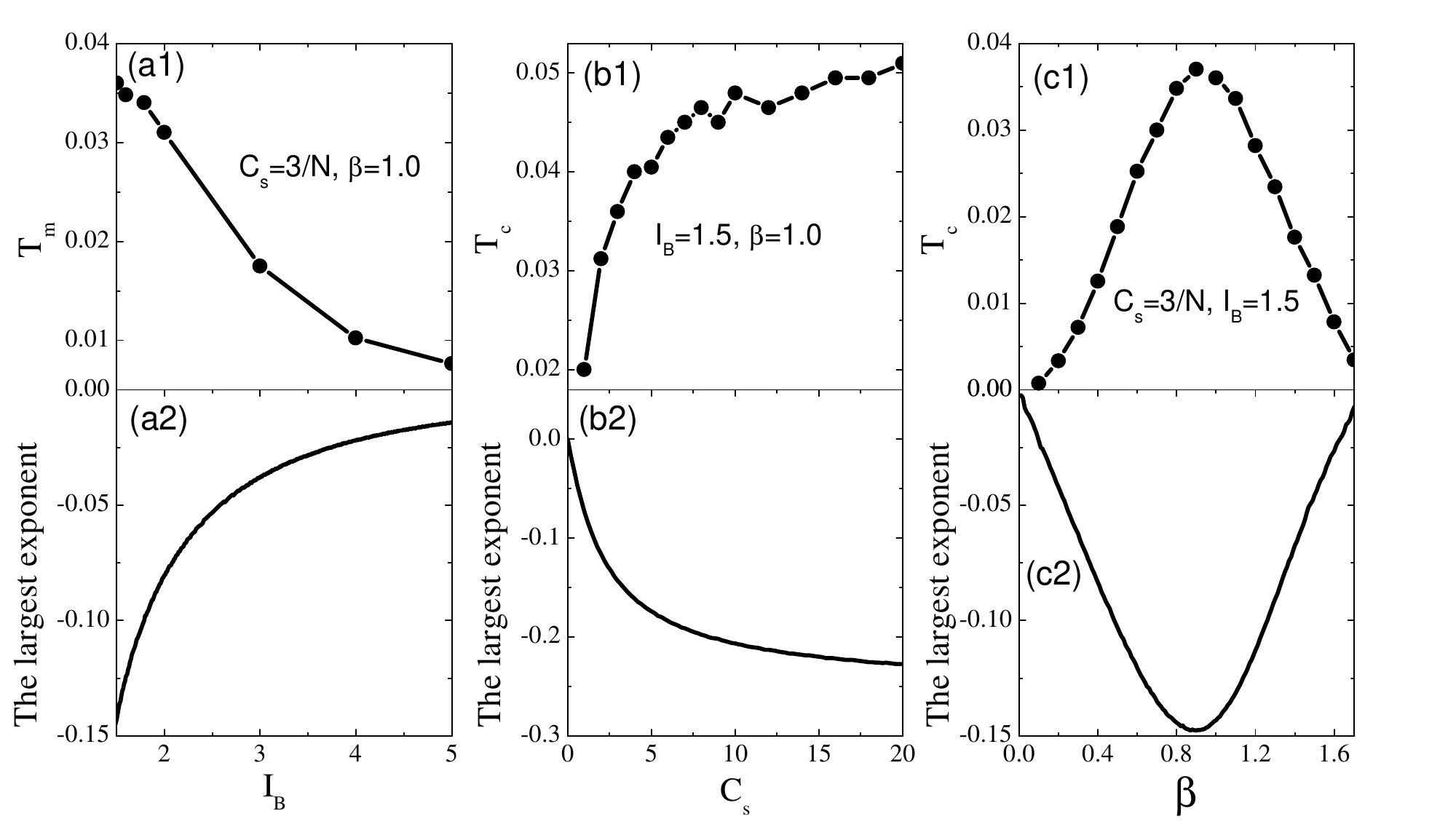,width=18cm} \caption{\label{f10}(Color online). (a1), (b1) and (c1): dependence of $T_m$ on the bias current, shunt capacitance and $\beta$ respectively. (a2), (b2) and (c2) are the corresponding largest Floquet exponent. Other parameters used are shown in the figure.}
\end{figure*}

Suppose the system initially at the uniform-oscillation state, and then we turn on thermal noise. The noise perturb the system away from the uniform state. However, the deviation from the attractor is penalized by the action $S$. The comparison of the action $S$ for the system moving from the attractor to the separatrix with the noise strength defines three distinct regions.
\begin{enumerate}
  \item $k_b T\ll S$, in this case the possibility of thermal escape is extremely small and this region is described by the reaction-rate theory\cite{Hanggi90}. Especially, when the thermal activation between two attractors $a$ and $b$ is asymmetric, that is the action $S_a \gg S_b$, the system will spend much longer time in the attractor $b$. This is the situation of retrapping from resistive state to zero-voltage state for small $\beta\ll1$ discussed before. When the bias current $I_B$ is close to the retrapping current $I_B-I_r\ll1$, the resistive state is about to lose stability. So the presence of weak noise will destabilize the resistive state and the system evolves into the zero-voltage one. On the other hand, the energy barrier for the system to transform from zero-voltage state to resistive one again is large when $I_B\ll I_c$, and the noise are not strong enough to promote such a transition. So the system remains zero-voltage. Thus the thermal noise increase the retrapping current\cite{Jacob82}.
  \item $k_b T\gg S$, in this region the thermal energy is large enough to kick the system off the attractor of the coherent oscillation, and the synchronization is destroyed. The temperature at which the synchronization is destroyed is the synchronization-desynchronization transition temperature $T_m$.
  \item for $T<T_m$, the uniform oscillations survive. The noise current excites perturbations and the system
frequently deviates from the attractor. The dynamics of the perturbations are described by Eq. (\ref{eq8}). These perturbations broaden the linewidth of the frequency spectrum. The linewidth at $\omega\gg 1$ can be estimated as follows. For $\omega\gg 1$, the \emph{IV} is linear, so the noise current $I^n$ induces a noise voltage $I^n/\beta$. From the ac Josephson relation $\partial_t \phi=2eV/\hbar$, one easily obtain that the linewidth increases linearly with $T$ for Gaussian white noise.
\end{enumerate}

\begin{figure}[b]
\psfig{figure=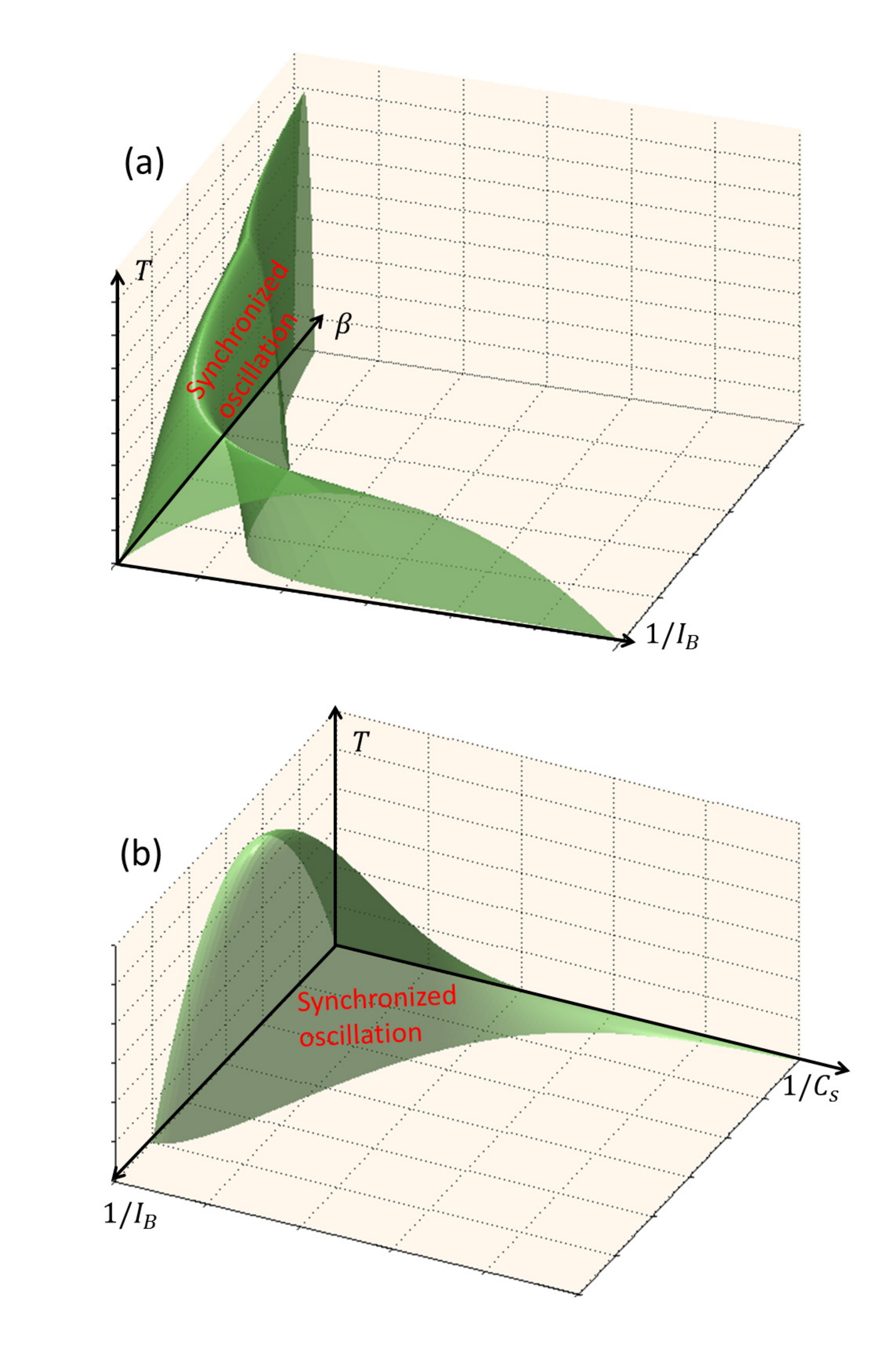,width=\columnwidth} \caption{\label{f11}(Color online). Possible stability diagram of the uniform solution (a) at a given $C_s$, and (b) at a given $\beta$. The region inside the green surface corresponds to stable uniform oscillations.}
\end{figure}

In the presence of noise, the equations of motion become stochastic, and it is natural to describe the dynamics in term of a probability density in the phase space. The flow of the probability density is governed by the Fokker-Planck equation. However analytical calculations of the coupled nonlinear partial differential Fokker-Planck equation is difficult. In this section we will use numerical simulations as a main workhorse, and we will also provide qualitative analysis to understand the numerical results. We first consider the desynchronization transition of the synchronous state and the critical behavior at the transition. We then find a correlation between the largest Floquet exponent and the transition temperature. Finally a stability diagram of the uniform oscillations with respect to noise is constructed.

\subsection{Synchronization-desynchronization transition}

To study the synchronization-desynchronization transition, we evaluate the order parameter defined in Eqs. (\ref{eqmelt1}) and (\ref{eqmelt2}), and its standard deviation
\begin{equation}\label{eqmelt3}
\sigma_r=\langle r^2 \rangle- \langle r \rangle^2,
\end{equation}
which is similar to the susceptibility defined in spin systems.

We solve numerically Eqs. (\ref{eq4}), (\ref{eq5}) and (\ref{eq6}), and derive $\langle r \rangle$ and
$\sigma_r$ at different $T$. The results are presented in Figs. \ref{f8} and \ref{f9}.
We also check the finite-size effect with different $N$'s. The finite size effect is prominent around $T_m$.
The synchronized oscillations is continuously suppressed by the thermal fluctuations.
At $T_m$ the synchronized oscillations become unstable, and the system undergoes a continuous transition into
random oscillations. Since  $\sigma_r$ serves as a measure of the fluctuation effect, it reaches maximum
at $T_m$, as shown in Fig. \ref{f9}. Practically $\sigma_r$ provides a convenient way to determine $T_m$
especially for a small system where the transition is obscured by the finite-size effect.

Once we identify the desynchronization transition as a critical phenomenon, we can define the exponent
\begin{equation}\label{eqmelt3e}
\langle r \rangle\sim (T_m-T)^{\beta_c}.
\end{equation}
In Fig. \ref{f8}, we obtain $\beta_c\approx1/2$ which is consistent with the mean-field theory.

We then study which factors determine $T_m$ and how to enhance $T_m$. As discussed at the beginning of this section, $T_m$ is given by the action for the system moving out of the attractor. Thus knowledge about the whole basin of attraction is needed. However, it is still conceivable that the local slope near the attractor may to certain extent reflect the global structure of basin. The local slope is just the largest Floquet exponent obtained in the previous section. Therefore one expects that the smaller the exponent, the higher $T_m$. We find numerically that it is indeed the case, as shown in Fig. \ref{f10}.

The correlation between the largest Floquet exponent and $T_m$ can be understood in terms of the local stability analysis. The small perturbations to the uniform oscillation decay $\widetilde{q}\sim \exp(\lambda_2 t)$ with $\lambda_2<0$ being the largest Floquet exponent. This is equivalent to the relaxation of a particle in the parabolic potential $\partial_t \widetilde{q}=-\partial V/\partial \widetilde{q}$ with $V(\widetilde{q})=-\lambda_2 \widetilde{q}^2/2$. The slope $-\lambda_2>0$ measures the depth of the potential. Thus it is more robust against noise for a larger $-\lambda_2$.

\subsection{Stability phase diagram with noise}
Based on the previous analysis, we discuss the stability phase diagram of the synchronization in the presence of thermal noise. For a given $C_s$, when the bias current is increased, the system approaches the synchronous state, where the associated Floquet exponent changes from positive to negative at the stability boundary. If the current increases further, it reach the maximal value $-\beta/2$. For a sufficient large current, the system becomes neutral stable according to Eq. (\ref{eq17}). Therefore the critical temperature first increases and then decreases with the current. The corresponding stability diagram is shown in Fig. \ref{f11}(a). Meanwhile, the shunt capacitance plays a role of coupling strength, so $T_m$ increases with $C_s$. Keep in mind that the current at the stability boundary is the retrapping current. At a given $T$, the retrapping current increase with $C_s$ for a small $\beta$ , while it decreases for a large $\beta$. Based on these observations, we construct the phase diagram of the coherent oscillations for a given $\beta$, which is sketched in Fig. \ref{f11}(b). The region enclosed with green surface represents a stable synchronization.

To enhance $T_m$ a larger $\beta$ is helpful since the maximal Floquet exponent is $-\beta/2$. One should also adjust the current accordingly to ensure that the maximum is reached. For a given operating frequency, one may increase $C_s$ to enhance the thermal stability, at sacrifice of the oscillating amplitude.

\section{Dynamic relaxation}

\begin{figure*}[t]
\psfig{figure=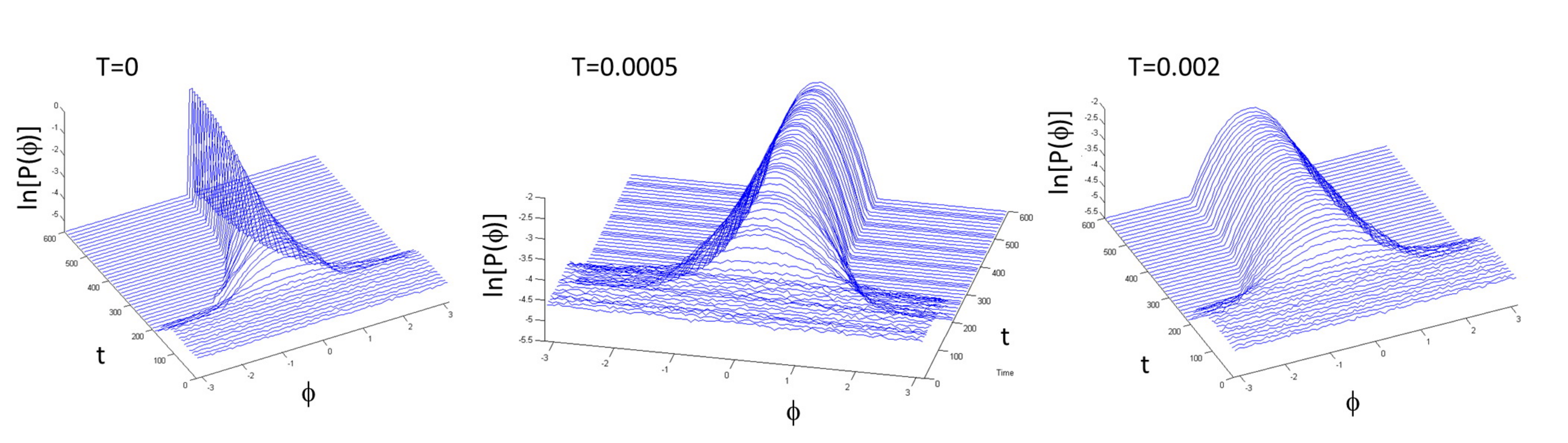,width=18cm} \caption{\label{f12}(Color online). Time evolution of the distribution of phase difference Eq. (\ref{eqdr0}), starting from completely random state at several temperatures. Here $\beta=0.02$, $I_B=1.5$ and $C_s=3/N$.}
\end{figure*}

\begin{figure}[b]
\psfig{figure=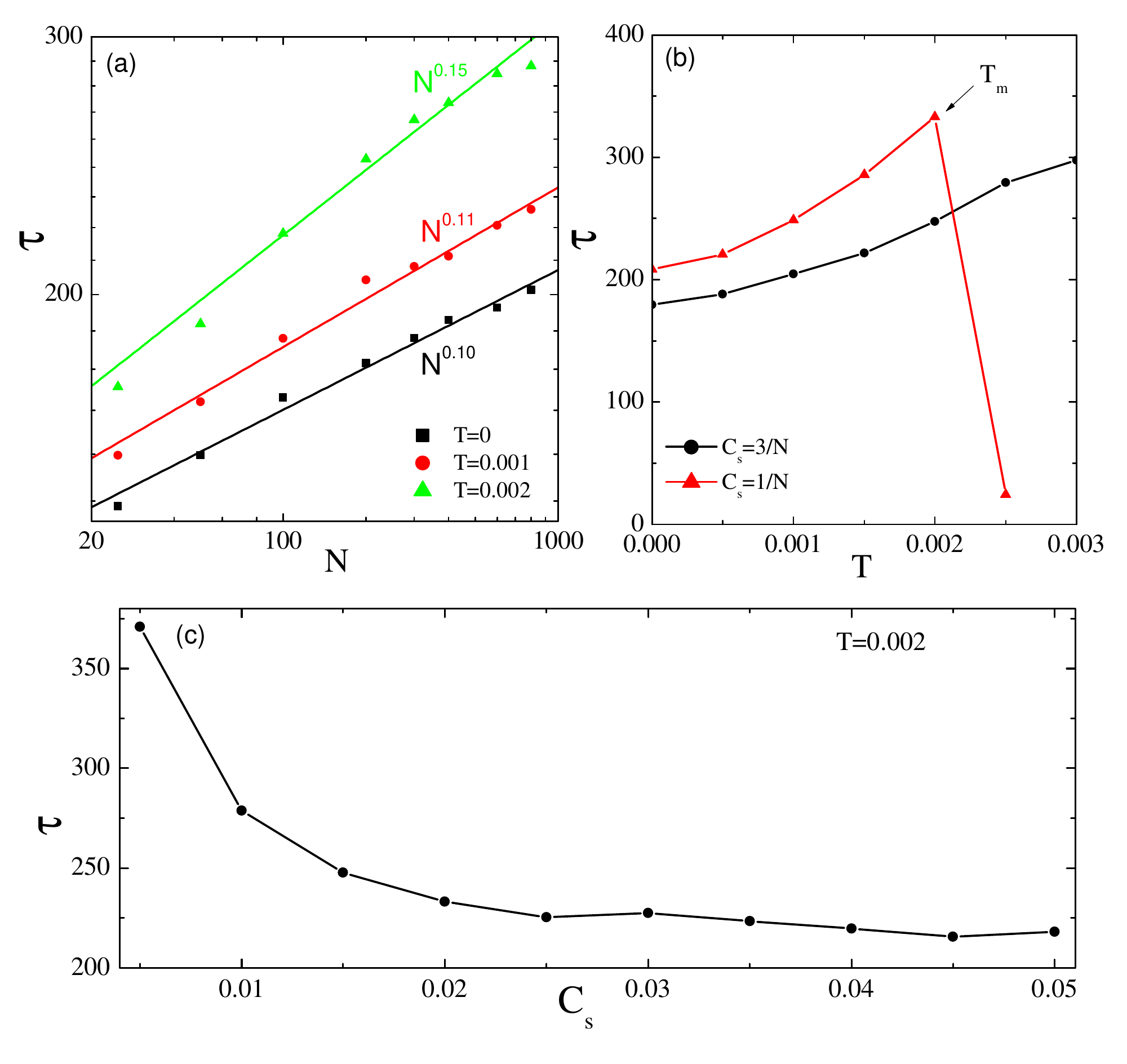,width=\columnwidth} \caption{\label{f13}(Color online). (a) Dependence of the relaxation time on the system size with different temperatures. Symbols are numerics and lines are the best fitting. (b) Dependence of the relaxation time on the temperature with the system size of $N=200$ and with different shunt capacitance. (c) Speedup of the relaxation by increasing the shunt capacitance. Here the system size is $N=200$ and temperature $T=0.002$. All these results are obtained with $\beta=0.02$ and $I_B=0.15$.}
\end{figure}

So far we have concentrated on the stability of the uniform oscillation, and investigated the dynamics of perturbations around the uniform state. However, in most applications, the initial condition cannot be guaranteed as the state of uniform oscillations. For instance, when we ramp up the current and bias all junctions in the resistive state, the initial state may be far away from the uniform state in the phase space. Therefore it is important to understand how the system approaches the uniform state. In the present study, we focus on the relaxation from disordered state (all junctions oscillate out-phase) into ordered state (all junctions oscillate uniformly). For a system whose final ordered state is in \emph{equilibrium}, this is a phase ordering phenomenon. The kinetics of phase ordering has been extensively studied decades ago in spin systems, and they can be described by universal scaling behavior\cite{Bray94}. However the relaxation dynamics is not very clear when the final state is \emph{out of equilibrium}.

To reach the uniform state, the initial state must be in the basin of attraction of the uniform state. This can be realized by operating all junctions at resistive state. We prepare the initial state with arbitrary nonzero $\langle r \rangle\ll1$. We also give initial velocity to all junctions, as such the system falls into the basin of attraction of the uniform state. Let us first consider dynamical relaxation obtained by computer simulation. We introduce the distribution of the phase difference between junctions
\begin{equation}\label{eqdr0}
P(\phi ) = \sum\limits_{i,j} {\delta (\phi  - \Delta_{ij})}
\end{equation}
with $\Delta_{ij} = {\phi _i} - {\phi _j}$. The time evolution of $P(\phi )$ is depicted in Fig. \ref{f12}. Initially the distribution is flat indicating a disordered phase. This flat distribution does not change with time too much at the beginning, but then it suddenly becomes sharp. Finally it reaches a steady distribution with finite width depending on the temperature.

Qualitative picture of the relaxation can be obtained based on the local stability analysis presented in Section III. Suppose we have a small synchronized cluster of junctions with population $N_i$ and the rest of junctions oscillate randomly. This small cluster serves as a seed of the nucleation, and deliver energy into the shunt capacitor, which in turn attracts nearby out-phase oscillators into the cluster. The growth rate of the synchronized population can be estimated by the local stability analysis by replacing $N$ with $n(t)$, where $n(t)$ is size of the cluster at time $t$. The time evolution of the population of the cluster follows
\begin{equation}\label{eqdr1}
n(t+d t) = n(t)\exp(-\lambda d t)\approx n(t) (1-\lambda(n) dt),
\end{equation}
where $dt$ is a small time step and $\lambda(n)<0$ is the largest Floquet exponent with cluster size of $n$. Then the time required for the system achieves global synchronization is
\begin{equation}\label{eqdr1b}
\tau_g = -\int_{N_i}^{N} dn \frac{1}{ n \lambda(n)}.
\end{equation}
Several observations are in order. First, the synchronization time $\tau_g$ increases with the total number of junctions $N$. Secondly, since $\lambda(n)<0$ decreases monotonically with $n$ and then saturate at $-\beta/2$ [see Eq. (\ref{eq17})], the initial relaxation is slow and it gradually speeds up, in accordance with Fig. \ref{f12}. Thirdly, in the presence of thermal fluctuations, thermal noise may kick oscillators out of the synchronized cluster. Thus the increase rate is reduced and relaxation time increases.

To quantify the relaxation process, we define the linear relaxation function\cite{LandauBook}
\begin{equation}\label{eqdr2}
A(t) = \frac{{\left\langle {r(t)} \right\rangle  - \left\langle {r(\infty )} \right\rangle }}{{\left\langle {r(0)} \right\rangle  - \left\langle {r(\infty )} \right\rangle }}.
\end{equation}
It starts from unity at $t=0$ and decays to $0$ in the steady state. The relaxation time is defined as
\begin{equation}\label{eqdr3}
\tau  = \int\limits_0^\infty  {A(t)dt}.
\end{equation}
For an exponential decay, the definition above is equivalent to the conventionally defined relaxation time. Two-stage relaxation for $A(t)$ is found at $T=0$. First $r$ increase from $0$ to a value close to $1$, where the local stability theory applies. Then the system relaxes into the ordered state exponentially with the exponent given by the Floquet exponents. Thermal fluctuations smear the distinction of the two-stage relaxation.

We numerically calculate $A(t)$ and compute $\tau$. The dependence of $\tau$ on the number of junctions $N$, temperature and shunt capacitor is plotted in Fig. \ref{f13}. $\tau$ increases with $N$ as expected from the qualitative estimate above. Furthermore the relaxation time follows a power law $\tau\sim N^{z'}$. The exponent $z'(T)$ increases with $T$. $\tau$ increases with temperature. At $T_m$ it diverges and then drops. (the relaxation time above $T_m$ is not very meaningful because the final state is also disordered.) Critical behavior is also identified for $\tau$ near $T_m$. On the other hand, $\tau$ decreases with $C_s$ which suggests a practical way to speed up the relaxation. This can be explained by regarding $C_s$ as a coupling strength of the system. A larger $C_s$ therefore increases the rigidity of the uniform solution.

\section{Conclusion}
In short, we have studied the synchronization of one dimensional array of point Josephson junctions coupled to a shunt capacitor. In the case of noise-free system, a stability phase diagram of the uniform oscillation is constructed. For strong damping, after the uniform solution becomes unstable, the system evolves into partially synchronized state. When the bias current is reduced below the Josephson critical current, the system becomes zero-voltage. For weak damping or moderate damping, after the instability of the uniform solution, the system evolves into the zero-voltage state. At transition the current is the experimentally measurable retrapping current. The retrapping current is increased by the shunt capacitor for weak damping ($\beta\lesssim 0.5$), while it decreases for moderate and strong damping ($\beta\gtrsim 0.5$). Thus transport measurement provides a convenient probe of the underlying dynamics. Similar results are obtained when a resistor is serially connected to the shunt capacitor.

In the presence of strong thermal noise, the coherent oscillation is destroyed through a second order phase transition. The critical exponent for the order parameter is $1/2$ in accordance with the mean-field theory. We also find the fluctuations of the order parameter $r$ showing a maximum at the transition, which may serve as a convenient quantity to locate the transition temperature. For a smaller relaxation time in the case of weak perturbations, the transition temperature is higher. The results suggest several possible ways to enhance the thermal stability.

The dynamic relaxation from a disordered phase to ordered state is then investigated. The relaxation time increases with the system size by a power law. It also increases when the system approaches the transition temperature from below. One may speed up the relaxation with a larger shunt capacitance.

Finally, a possible phase diagram of the uniform solution is proposed when thermal fluctuations are involved. Our results are of importance for the design of useful superconducting devices based on Josephson junctions arrays.

\section{Acknowledgement}

SZL and XH are supported by WPI Initiative on Materials Nanoarchitronics, MEXT, Japan and CREST-JST Japan. LB is supported by the National Nuclear Security Administration of the U.S. Department of Energy at Los Alamos National Laboratory under Contract No. DE-AC52-06NA25396 and by the LANL/LDRD Program.


\end{document}